\documentclass[authoryear,12pt,letter,3p]{jowarticle}
\usepackage{graphicx}
\usepackage{amsmath}
\usepackage{amssymb}
\usepackage{setspace}
\usepackage[all]{xy}
\makeatletter
\renewcommand\section{\@startsection{section}{1}{\z@}{-3.25ex plus -1ex minus -.2ex}{1.5ex plus .2ex}{\normalsize\bf}}
\renewcommand\subsection{\@startsection{subsection}{2}{\z@}{-3.25ex plus -1ex minus -.2ex}{1.5ex plus .2ex}{\normalsize\bf}}
\renewcommand\subsubsection{\@startsection{subsubsection}{3}{\z@}{-3.25ex plus -1ex minus -.2ex}{1.5ex plus .2ex}{\normalsize\bf}}
\makeatother

\begin{document}
\begin{frontmatter}
\title{Regarding the ``Hole Argument''}
\author{James Owen Weatherall}\ead{weatherj@uci.edu}
\address{Department of Logic and Philosophy of Science\\ University of California, Irvine, CA 92697}
\begin{abstract}I argue that the Hole Argument is based on a misleading use of the mathematical formalism of general relativity.  If one is attentive to mathematical practice, I will argue, the Hole Argument is blocked.\end{abstract}
\end{frontmatter}

\doublespacing

\section{Introduction}\label{introduction}

Few topics in the foundations of physics have received more attention in the last thirty years than Einstein's ``Hole Argument''.\footnote{For more on the Hole Argument, see \citet{NortonSEP} and references therein.  For an up-to-date overview of debates concerning relationism and substantivalism and the role the Hole Argument has played, see \citet{PooleyOHPP}.  See also \citet{EarmanWEST}.}  When it first occurred to him in 1913, Einstein feared that the Hole Argument ruled out any ``generally covariant'' theory of gravitation.  But by 1915, he had rejected the argument, convinced that he had discovered a gravitational theory of just the sort he desired.  For the next sixty five years, the Hole Argument was seen as a historical curiosity, little more than a misstep on the way to general relativity.  This changed when \citet{Earman+Norton} resurrected the argument, now with spacetime substantivalism as its target.\footnote{\citet{NortonSEP} credits \citet{Stachel} as the first person to recognize the ``non-triviality'' of the Hole Argument.  (Stachel's paper was first read at a conference in 1980, but only published later.)  See also \citet{Norton1984}, \citet{Stachel1986}, and \citet{Earman1986} for preludes to \citet{Earman+Norton}.}  It is in this latter form that the argument has attracted so much attention from philosophers and foundationally-oriented physicists.

The basic thesis of the present note is that Einstein and the generations of physicists and mathematicians after him were right to reject the Hole Argument.  It is based on a misleading use of the mathematical formalism of general relativity.  If one is attentive to mathematical practice, I will argue, the Hole Argument is blocked.

This thesis may sound radical, so let me be clear about just what I intend to say.  As presented by \citet{Earman+Norton}, the Hole Argument concerns two metaphysical doctrines: one is something that they call ``Leibniz equivalence''; the other is a certain variety of indeterminism.  Their claim is that if one rejects Leibniz equivalence in relativity theory, one is compelled to accept this particular form of indeterminism.  In this form, one might think of the Hole Argument as a challenge for interpreters of general relativity---particularly for a certain stripe of ``substantivalist'', but also (perhaps) more generally.  The challenge is to find a way of interpreting the formalism so as to avoid this form of indeterminism.  The present paper may then be understood as providing one option for how to meet this challenge.  This particular option is distinctive---and, I think, attractive---because it is essentially neutral on the metaphysics of space and time.\footnote{When I say ``neutral,'' I mean only that the arguments I give here do not compel one to accept any particular view on the metaphysics of space and time, which is not to say that the arguments of the present paper could not be put into quick service in such debates. I will elaborate this point in section \ref{revisited}.}  Indeed, for the bulk of the paper, I will set aside the interpretation of the Hole Argument as concerning Leibniz equivalence and indeterminism (or substantivalism and relationism). My focus will be on the mathematics underlying the Hole Argument.\footnote{\label{formal}This strategy may remind some readers of the ``syntactic'' or ``formal'' responses to the Hole Argument found in \citet{Mundy} and \citet{Leeds}. (See \citet{RynasiewiczSyntax} for a reply to these.)  But I would resist this way of understanding what I am doing.  Instead, I take myself to be making a conceptual point about the way in which mathematics is used in physics.  In that vein, this paper's closest cousin may be \citet[esp. \S XVI]{Wilson}.  (In another sense, the arguments here are close to those of \citet{Butterfield} and \citet{Brighouse}, in ways I will expand on in section \ref{revisited}.)}

This strategy may immediately raise a worry.  The Hole Argument concerns determinism and the discrimination of possible worlds.  An argument that focuses exclusively on mathematics cannot address what is ultimately an interpretive, or even metaphysical, question about when two possible worlds are distinct.\footnote{\label{possibleWorlds} A version of this point was recently emphasized by \citet[pg. 126]{Dasgupta}.}  I am happy to concede this point, with two caveats.  First, (1) our interpretations of our physical theories should be guided by the formalism of those theories.  And second, (2) insofar as they are so guided, we need to be sure that we are using the formalism correctly, consistently, and according to our best understanding of the mathematics.  My argument here concerns only (2).  In particular, I do not intend to argue that there is a mathematical solution to an interpretational problem.  Rather, I will argue that the mathematical argument that allegedly generates the interpretational problem is misleading.  The moral is that the standard formalism of relativity theory does not \emph{force} us to confront a dilemma about indeterminism.\footnote{Nor does it force us to change the formalism of relativity theory to avoid such a dilemma, as has been proposed, for instance, by \citet{Earman1986, EarmanWEST} and \citet{Iftime+Stachel}.  (For more on Earman's proposal in particular, which concerns the formalism of Einstein algebras \citep{GerochEA}, see \citet{RynasiewiczEA} and \citet{Rosenstock+etal}.)  This upshot is particularly important in the context of other areas of philosophy of physics where the Hole Argument is invoked, such as the literature on gauge theories and Yang-Mills theory.  See \citet{Healey}, \citet{Arntzenius}, \citet{WeatherallGauge}, \citet{WeatherallYM}, and \citet{Rosenstock+Weatherall1}.}

The two caveats in the previous paragraph require some unpacking.  In particular, regarding (1), there are many ways in which one might expect a mathematical formalism to ``guide'' interpretation, and just as many ways in which such ``guidance'' might go awry.  I do not claim that there is some straightforward way to move from formalism to interpretation.  For the purposes of the present paper, the form of guidance I require is just this: the default sense of ``sameness'' or ``equivalence'' of mathematical models in physics should be the sense of equivalence given by the mathematics used in formulating those models.

This brings us to (2).  In contemporary mathematics, the relevant standard of sameness for mathematical objects of a given kind is given by the mathematical theory of those objects.  In most cases, the standard of sameness for mathematical objects is some form of isomorphism.\footnote{Here and throughout, I am using the term ``isomorphism'' in the broad sense of category theory, as any invertible map within a category.  This means that maps that are standardly called isomorphisms, such as ring isomorphisms or group isomorphisms, count as isomorphisms.  But other maps, such as diffeomorphisms between smooth manifolds or homeomorphisms between topological spaces, also count as isomorphisms in this sense.  Still, not all maps corresponding to a standard of ``sameness'' are isomorphisms.  For one example where the relevant standard is usually \emph{not} isomorphism, consider categories.  Categories are standardly taken to be ``the same'' when they are equivalent, where equivalence of categories is a weaker notion than isomorphism of categories.  But here, too, this is a feature of the relevant mathematical theory.  For more on categories, see \citet{MacLane}, \citet{Borceux}, or \citet{Awodey}.}  Familiar mathematical objects---groups, manifolds, graphs---are only defined up to isomorphism.  This means that two isomorphic groups (say) may be distinct in all sorts of ways---they may be constructed from different sets, their operations may be defined differently, they may be associated with different rings---but there is one way in which they are the same: they are the same \emph{as groups}, or in other words, they have the same group structure.\footnote{There is a sense in which the view described here is a variety of mathematical structuralism \citep{Shapiro}.  But I want to be cautious about this term.  Structuralism is often (though not always) associated with a view about the ontology of mathematics, on which ``structures'' are taken to have some robust ontological status.  I do not need to endorse any such view here.  Similarly, I take it that the views on structure in mathematics defended by \citet{MaddyDtA} and \citet{Burgess} are compatible with everything I say here, and neither Maddy nor Burgess would take on the mantle of structuralism.}

Taken together, then, at least for present purposes, the caveats amount to the view that mathematical models of a physical theory are only defined up to isomorphism, where the standard of isomorphism is given by the mathematical theory of whatever mathematical objects the theory takes as its models.  One consequence of this view is that isomorphic mathematical models in physics should be taken to have the same representational capacities.  By this I mean that if a particular mathematical model may be used to represent a given physical situation, then any isomorphic model may be used to represent that situation equally well.\footnote{To explain this way of putting the point it may be worth being explicit, here, that I take mathematical models to be devices that we use for representational purposes, not objects that stand in some context- or use-independent relation to physical situations that they ``represent''.  How we use mathematical models to represent physical situations is something we get to stipulate in practice.  When I say in the sequel that such and such model ``represents'' some situation, it should be taken as short-hand for ``may be used to represent (by the lights of the relevant theory)''.}  Note that this does not commit me to the view that equivalence classes of isomorphic models are somehow in one-to-one correspondence with distinct physical situations.\footnote{See \citet{Belot} for arguments against this view.  Conversely, taking the \emph{members} of an equivalence class of isomorphic models to be in one-to-one correspondence with distinct physical possibilities is just as troublesome, since then differences in how we construct mathematical objects from sets (as highlighted by, say, \citet{Benacerraf}) would lead to a proliferation of physical possibilities.}  But it does imply that if two isomorphic models may be used to represent two distinct physical situations, then each of those models individually may be used to represent \emph{both} situations.

The views expressed in the previous paragraphs seem reasonable to me, though I recognize that they may be controversial.  Still, I will not to defend them here.  Rather, I take them as background commitments that inform the arguments to follow.  Some readers, then, will want to take what follows in conditional form: if one is committed to the views about (applied) mathematics just described, then (I claim) the reaction to the Hole Argument described here follows.  These commitments should also clarify the strategy described above, of setting aside the Earman-Norton interpretation of the Hole Argument to focus on the mathematical argument.  My arguments below will not be devoid of interpretation of any sort.  But the interpretational issues I will address will concern the applied mathematics.  What I take myself to be setting aside, at least initially, is the further interpretational issue of how the Hole Argument bears on issues of determinism or the ontology of spacetime.

The plan from here will be as follows.  I will begin with a warm-up exercise.  I will then reconstruct the Hole Argument and describe what I take to be the problem.  Next, I will contrast the Hole Argument with a superficially similar argument in the context of classical spacetime theory, with the goal of further illuminating what makes the Hole Argument misleading.  I will conclude by returning to the Hole Argument and addressing how one might understand the morals often drawn from it in light of the arguments of the present paper.

\section{A Warm-Up Exercise}\label{warm-up}

Consider the following argument.  We begin with the set of integers, $\mathbb{Z}$.\footnote{I do not believe there is any ambiguity regarding what the ``set of integers'' is supposed to be.  But for the sake of precision, I take it to be a countably infinite set, along with canonical (and exhaustive) labels for the elements.  To this, one might add order structure, group structure, etc. in canonical ways.  I do not have a particular \emph{construction} of the integers in mind, and for present purposes, nothing trades on the details of any particular construction.  For more on this last point, see \citet{Benacerraf} and \citet{Burgess}.}   These are naturally understood to form a group under addition, where $0$ is the additive identity.  I will denote this group by $(\mathbb{Z},+,0)$.  Now consider the following map, $\varphi:\mathbb{Z}\rightarrow\mathbb{Z}$, defined by $\varphi:n\mapsto n+1$ for every $n\in\mathbb{Z}$.  This map is evidently bijective, and thus invertible.  It can also be used to induce a group isomorphism, as follows.  We define a group whose underlying set is the set $\mathbb{Z}$, but whose group structure is inherited from $\varphi$.  That is, the identity of this group is $\tilde{0}=\varphi(0)=1$; for any $n\in\mathbb{Z}$, the inverse of $n$ is given by $n^{\widetilde{-1}}=2-n$; and for any $n,m\in\mathbb{Z}$, addition is given by $n \tilde{+} m= n+ m-1$.  This group will be denoted by $(\mathbb{Z},\tilde{+},\tilde{0})$.  Under this definition, one can easily confirm that for any $n,m\in \mathbb{Z}$, $\varphi(n - m)=\varphi(n)\tilde{+}\varphi(m)^{\widetilde{-1}}$,\footnote{To see this, note that for any $n,m\in\mathbb{Z}$, $\varphi(n)\tilde{+}\varphi(m)^{\widetilde{-1}}=(n+1)\tilde{+}(m+1)^{\widetilde{-1}}=(n+1)\tilde{+}(2-(m+1))=(n+1)\tilde{+}(1-m)=(n+1)+(1-m)-1=n-m+1 = \varphi(n-m)$.} which, since $\varphi$ is bijective, is sufficient to show that $\varphi$ is a group isomorphism.

So far so good.  But now (so the argument goes) we face a puzzle.  The sets underlying $(\mathbb{Z},+,0)$ and $(\mathbb{Z},\tilde{+},\tilde{0})$ are identical: they have precisely the same elements.  And as groups, they are isomorphic, which means that they are fully equivalent with regard to their group structure.  Indeed, there is a sense in which both are just the group of integers under addition.\footnote{One might make this claim precise by characterizing the group of integers by its group theoretic properties.  For instance, the group of integers is the unique (up to isomorphism) free group on a singleton set.  One can easily confirm that $(\mathbb{Z},+,0)$ and $(\mathbb{Z},\tilde{+},\tilde{0})$ are both free groups on one element.  Note, too, that the ``canonical labels'' associated with the integers play no role in the group structure.  This is clear because, as the example illustrates, these labels are not preserved under group isomorphisms.}  But they disagree about what the identity element is.  In $(\mathbb{Z},+,0)$, the identity is $0$; in $(\mathbb{Z},\tilde{+},\tilde{0})$, it is $\tilde{0}=1$.  And so it would appear that there is an ambiguity with regard to which number is the identity in the group of integers.

This argument should give one pause.  There is no ambiguity regarding the additive identity of the integers.  Indeed, the identity element is provably unique for any group.  Here, we have simply introduced a trivial notational variation, where we assign the symbols ``$\tilde{0}$'' or ``$1$'' to the identity in $(\mathbb{Z},\tilde{+},\tilde{0})$ and the symbol ``$0$'' to the identity in $(\mathbb{Z},+,0)$.  We know this notational variation does not affect the group structure, since we have exhibited a group isomorphism between $(\mathbb{Z},+,0)$ and $(\mathbb{Z},\tilde{+},\tilde{0})$.  So it would seem something has gone wrong in the previous argument.  One way of putting the problem is that the argument conflates two different senses of equivalence.  When I defined $(\mathbb{Z},\tilde{+},\tilde{0})$, I stipulated that it was a group ``whose underlying set is the set $\mathbb{Z}$''.  In this way, I implicitly defined a bijective map $1_{\mathbb{Z}}:\mathbb{Z}\rightarrow\mathbb{Z}$, where for any $n\in\mathbb{Z}$, $1_{\mathbb{Z}}:n\mapsto n$. This is the identity map on the underlying sets of the two groups.  It is (again implicit) reference to this map that justifies the claim that, as sets, $(\mathbb{Z},+,0)$ is ``identical to'' $(\mathbb{Z},\tilde{+},\tilde{0})$.

I then went on to define the group structure of $(\mathbb{Z},\tilde{+},\tilde{0})$ in such a way that $(\mathbb{Z},+,0)$ and $(\mathbb{Z},\tilde{+},\tilde{0})$ were isomorphic by construction.  This isomorphism was realized by the map $\tilde{\varphi}:(\mathbb{Z},+,0)\rightarrow(\mathbb{Z},\tilde{+},\tilde{0})$.  (I have written $\tilde{\varphi}$ here because the map $\varphi$ used to induce the isomorphism had a different codomain---namely $\mathbb{Z}$.)  And so the claim that $(\mathbb{Z},+,0)$ and $(\mathbb{Z},\tilde{+},\tilde{0})$ are isomorphic---that is, they ``have the same group structure''---is made with reference to $\tilde{\varphi}$.  But of course, $\tilde{\varphi}$ and $1_{\mathbb{Z}}$ are different maps.  And $1_{\mathbb{Z}}$ is not, nor does it induce, an isomorphism between the groups $(\mathbb{Z},+,0)$ and $(\mathbb{Z},\tilde{+},\tilde{0})$.

The problem, then, comes in using $1_{\mathbb{Z}}$ to compare the identity elements of the isomorphic groups $(\mathbb{Z},+,0)$ and $(\mathbb{Z},\tilde{+},\tilde{0})$.  Since $1_{\mathbb{Z}}$ does not realize the isomorphism, it is not the relevant standard of comparison for the present purpose; conversely, if one insists on comparing $(\mathbb{Z},+,0)$ and $(\mathbb{Z},\tilde{+},\tilde{0})$ using $1_{\mathbb{Z}}$, one loses the grounds on which one claimed the groups were equivalent in the first place.  The general moral may be put as follows: all assertions of relations between mathematical objects---including isomorphism, identity, inclusion, etc.---are made relative to some choice of map.  In many cases, one implicitly uses the identity map to describe such relations.  But, as the example above should make clear, one has to be very careful when one does so. Often the map realizing a relation is not the identity at all, in which case the identity map is not the relevant standard of comparison.\footnote{Of course, there are \emph{some} senses in which the identity map is privileged, mathematically speaking.  For instance, given any mathematical object $A$, the identity map $1_A:A\rightarrow A$ is the unique map such that, given any other map $\zeta:A\rightarrow A$, $\zeta\circ 1_A = 1_A\circ \zeta = \zeta$. But the fact that the identity has certain mathematical properties that distinguish it from other maps does not mean that it is privileged for the purposes of characterizing relationships between mathematical objects.}

The moral, here, is that the assertion above, that ``[the two groups] disagree about what the identity [of the group of integers] is,'' was a non-sequitur.   This is because the claim being made amounts to the observation that $0=1_{\mathbb{Z}}(0)\neq \tilde{0}=1$, and here one is (incorrectly) using the map $1_{\mathbb{Z}}$ to compare elements of $(\mathbb{Z},+,0)$ and $(\mathbb{Z},\tilde{+},\tilde{0})$.  If one refers only to the map $\tilde{\varphi}$ that realizes the isomorphism, the apparent ambiguity does not arise.  One has, as required, $\varphi(0)=1=\tilde{0}$.

\section{The Hole Argument}\label{main-event}

With the example of the previous section in mind, we can now turn to the Hole Argument.  At the heart of the Hole Argument is a mathematical argument that is strikingly similar to the one I just discussed.

This argument may be reconstructed as follows.  Fix a model of a relativity theory, which is a relativistic spacetime, i.e., a Lorentzian manifold, $(M,g_{ab})$.\footnote{\citet{Earman+Norton} seem to take for granted that the models of relativity theory are manifolds with various fields defined on them.  Supposing that a Lorentzian manifold $(M,g_{ab})$ is a solution to Einstein's equation allows us to reconstruct what the energy-momentum content of the spacetime must be, and so stipulating just the manifold and metric is sufficient to specify the model.  (That said, fixing the energy-momentum tensor will not, in general, be sufficient to fix the physical matter fields giving rise to that energy-momentum tensor.  But everything I say here could be carried over [\emph{mutatis mutandis}] to the context where one stipulates additional fields on spacetime, so I will focus on energy-momentum here without loss of generality.)  In what follows, I use the terms ``Lorentzian manifold'' and ``relativistic spacetime'' interchangeably.}  Now consider a diffeomorphism (i.e., a smooth map with smooth inverse) $\psi:M\rightarrow M$ with the following properties: there exists some (proper) open set $O\subset M$, with compact closure, such that $\psi$ acts as the identity on $M-O$, but on $O$, $\psi$ is \emph{not} the identity.  This is sufficient to guarantee that $\psi$ is not the identity map on $M$.  We can then use $\psi$ to induce an isometry as follows.  We define a relativistic spacetime, $(M,\tilde{g}_{ab})$, whose underlying manifold is once again $M$, and whose metric is defined by $\tilde{g}_{ab}=\psi_*(g_{ab})$, where $\psi_*$ is the pushforward map determined by $\psi$.\footnote{Recall that in the case of diffeomorphisms, one can define pushforward maps for tensorial objects with both contravariant and covariant indices.}  One can easily confirm that under this definition, $(M,g_{ab})$ and $(M,\tilde{g}_{ab})$ are isometric spacetimes, with the isometry realized by $\psi$.  Now observe that, according to the theory of smooth manifolds, diffeomorphism is the standard of isomorphism for manifolds; just as other mathematical objects are only defined up to isomorphism, manifolds are only defined up to diffeomorphism.\footnote{See \citet[pg. 5]{ONeill} and \citet[pg. 36]{Lee} for explicit statements to this effect in the geometry literature.}  Isometry, meanwhile, is the standard of isomorphism for (pseudo-)Riemannian manifolds, including Lorentzian manifolds.  So $(M,g_{ab})$ and $(M,\tilde{g}_{ab})$ are isomorphic as Lorentzian manifolds.

But now we have a problem (according to the Hole Argument).  These two spacetimes have the same underlying manifold.  And moreover, they are isometric, which means they have the same structure as Lorentzian manifolds, and thus (as Earman and Norton argue) they agree on all invariant, observable structure.\footnote{Here I am using Earman and Norton's language.  To put the claim in the language of sections \ref{introduction} and \ref{revisited}: since these spacetimes are isometric, they have the capacity to represent exactly the same physical situations.}  And yet, within the region $O$, \emph{there are points at which these two models disagree about the value of the metric}.  This suggests that there is an ambiguity with regard to the value of the metric at a point $p\in O$ at which the two metrics disagree. Returning, for a moment, to the intended interpretation of the argument, it is this ambiguity that Earman and Norton take to represent a kind of indeterminism.  The idea is that no matter what constraints one places on the metric outside of $O$ (including, say, at all times prior to $O$, provided the spacetime has a global time function), one cannot uniquely determine the metric within $O$.  And one is forced to accept indeterminism in this sense, they argue, unless one accepts---as a substantive additional assumption---Leibniz equivalence, which is the principle that ``diffeomorphic models are equivalent'' \citep[pg. 522]{Earman+Norton}.\footnote{\label{confusion} I think this way of expressing the principle may be a source of confusion, for reasons I will elaborate in fn. \ref{diffeomorphism}.}

It is at this point that I want to object.  I claim that the structure of the argument just presented is the same as the one I analyzed in the previous section.  On the one hand, there is the claim that the spacetimes $(M,g_{ab})$ and $(M,\tilde{g}_{ab})$ have the same ``underlying manifold'', just as the groups $(\mathbb{Z},+,0)$ and $(\mathbb{Z},\tilde{+},\tilde{0})$ have the same ``underlying set''.  Assertions concerning \emph{the} point $p$ in the context of both $(M,g_{ab})$ and $(M,\tilde{g}_{ab})$ are implicitly made relative to the identity map on this manifold, $1_M:M\rightarrow M$.  Meanwhile, we have the map $\psi:M\rightarrow M$, which is a diffeomorphism.  It is $\psi$ that gives rise to the isometry $\tilde{\psi}:(M,g_{ab})\rightarrow (M,\tilde{g}_{ab})$, where I have again used a $\tilde{\;}$ to be clear that $\psi$ and $\tilde{\psi}$ are different maps---$\psi$ is an automorphism of $M$, whereas $\tilde{\psi}$ is an isomorphism, but not an automorphism, between the Lorentzian manifolds $(M,g_{ab})$ and $(M,\tilde{g}_{ab})$.

When we say that $(M,g_{ab})$ and $(M,\tilde{g}_{ab})$ are isometric spacetimes, and thus that they have all of the same invariant, observable structure, we are comparing them relative to $\tilde{\psi}$.  Indeed, we must be, because as in the previous example, there is no sense in which $1_M$ either is or gives rise to an isometry.  In other words, relative to $1_M$, $(M,g_{ab})$ and $(M,\tilde{g}_{ab})$ are \emph{not} equivalent, physically or otherwise.  The reason is that there exist points $p\in O$ at which $(g_{ab})_{|p}\neq (\tilde{g}_{ab})_{|1_M(p)}$.  Consider an observer sitting at a point of spacetime represented by the point $p\in O$ of $(M,g_{ab})$.  If one were to attempt to represent that same observer's location by the point $1_M(p)=p$ of $(M,\tilde{g}_{ab})$, one would conclude that the observer would see measurably different metrical properties, different curvature, etc. In general only one of these assignments can be correct.\footnote{To be clear, the idea is that one can perfectly well consider two fields, $g_{ab}$ and $\tilde{g}_{ab}$ on the manifold $M$ such that $\varphi_*(g_{ab})=\tilde{g}_{ab}$.  But if one has an observer at a given point $p$, the situation where the metric at $p$ is $g_{ab}$ and the metric at that point is $\tilde{g}_{ab}$ will in general be distinguishable---for instance, in one case one might be happily working at one's desk, and in the other plummeting into a black hole.  (See also \citet{Maudlin1993} for a related point, regarding the epistemological significance of shift arguments.)}   Meanwhile, if one only considers $\tilde{\psi}$, no disagreement arises regarding the value of the metric at any given point, since for any point $p\in M$, $(g_{ab})_{|p}=(\tilde{g}_{ab})_{|\psi(p)}$ by construction.\footnote{This is not to say that the metric \emph{is} determined (up to isometry) within an arbitrary region $O$ by a full specification outside that region, even once one assumes Einstein's equation.  But whether or not it is fully determined will depend on features of the spacetime in question such as whether there exists an appropriate Cauchy surface.  That is to say, there are still senses in which general relativity may be indeterministic, but one does not have the kind of generic indeterminism associated with the Hole Argument.  Still, for many cases of interest, it is known that there \emph{do} exist unique (up to isometry) solutions to Einstein's equation.}

It might be helpful to make the point in the previous paragraph in a different way.  To begin, one might reason as follows.  Suppose we do compare the points of $(M,g_{ab})$ and $(M,\tilde{g}_{ab})$ using $1_M$.  As we have seen, by this standard, the two Lorentzian manifolds are not mathematically equivalent, in the sense that this map does not realize or induce an isometry.  Moreover, an observer at any particular spacetime point represented by a point $p\in M$ will be able to distinguish the case where the metric at $p$ is $(g_{ab})_{|p}$ from the case where the metric is $(\tilde{g}_{ab})_{|p}$, and so in this sense, if one compares the manifolds using $1_M$, the two spacetimes make different predictions, only one of which can be correct.  But, one might think, there is still a (different) sense in which $(M,g_{ab})$ and $(M,\tilde{g}_{ab})$ are equivalent: namely, they are observationally or empirically equivalent.  The idea here would be that for any region $R$ of spacetime that may be adequately represented by some region $U$ of $(M,g_{ab})$, there is a corresponding region $\tilde{U}=\psi[U]$ of $(M,\tilde{g}_{ab})$ that can represent that same region of spacetime equally well, for all purposes.  By the standard set by $1_M$, however, $U$ and $\tilde{U}$ correspond to distinct collections of points in $M$.  And so, we have a sense in which $(M,g_{ab})$ and $(M,\tilde{g}_{ab})$ are inequivalent as mathematical objects---again by the standard set by $1_M$---but nonetheless may represent the exact same phenomena, and thus, cannot be distinguished by any experiment.

This argument is just a recapitulation of the Hole Argument in different terms.  There are several remarks to make.  The first is that this new, different sense in which $(M,g_{ab})$ and $(M,\tilde{g}_{ab})$ turn out to be equivalent is precisely the sense we began with, namely that they are isometric.  One can see this by noting that, in the argument just given, the region $R$ of spacetime is represented by region $U$ of $(M,g_{ab})$ and by region $\tilde{U}=\psi[U]$ of $(M,\tilde{g}_{ab})$.  These regions are related by the diffeomorphism $\psi$ that generates the isometry between the spacetimes.  So even on this account, all one has done is argue that the spacetimes are empirically equivalent by appealing to the fact that they are mathematically equivalent as Lorentzian manifolds---namely, that they are related by $\tilde{\psi}$.  Indeed, the fact that such an isometry exists provides the \emph{only} sense in which the two spacetimes are empirically equivalent---just as the isometry provides the only sense in which the spacetimes are mathematically equivalent, and the group isomorphism described above provides the only sense in which the groups discussed in the previous section have the same structure as groups.

But we began by stipulating that we would compare points using $1_M$.  What should we make of this stipulation in light of the foregoing?  In a sense, we have not kept to it, since as we have just seen, the argument above turned on a relationship between the models that only holds relative to $\tilde{\psi}$---and indeed, if one is careful about \emph{only} relating points with $1_M$, the spacetimes are not empirically equivalent at all.  

Still, one might claim that, nonetheless, $1_M$ plays some role.  I see two ways of understanding such a claim.  On the first interpretation, one implicitly claims that the only map that \emph{could} preserve all of the salient structure of $(M,g_{ab})$ is the identity map.  This claim effectively denies that isometric spacetimes are mathematically equivalent---and thus, at least in the presence of the background views on mathematics described above, implies that one is not representing spacetime by (only) a Lorentzian manifold.  Instead, one must be invoking some other, additional structure that is not preserved by isometry and that allows us to draw a distinction between the way in which $(M,g_{ab})$ represents the world and the way in which $(M,\tilde{g}_{ab})$ does so.  Let me set this possibility aside for now, as it amounts to rejecting the assumption we began with, that we could adequately represent spacetime using (only) a Lorentzian manifold; I will return to to this proposal in section \ref{revisited}.

On the second interpretation, meanwhile, one maintains that even though isometry \emph{is} the relevant standard of sameness of the structures in question, and even though it is the existence of an isometry that licenses the claim that $(M,g_{ab})$ and $(M,\tilde{g}_{ab})$ are empirically equivalent, nonetheless $1_M$ may (or should) be understood to play a substantive role in relating the two spacetimes, even in the context of assertions of their empirical equivalence (which, as we have seen, is given by $\tilde{\psi}$).  But it is precisely this position that I claim is untenable.  The reason is that it runs together two distinct ways of relating $(M,g_{ab})$ and $(M,\tilde{g}_{ab})$.  And as we saw in the previous section, if one is not attentive to which maps are being used to establish a given relationship between two mathematical objects, one can draw very misleading conclusions.  Indeed, this is precisely what has happened in the Hole Argument.

This discussion may be summed up as follows. There is a sense in which $(M,g_{ab})$ and $(M,\tilde{g}_{ab})$ are the same, and there is a sense in which they are different.  The sense in which they are the same---that they are isometric, or isomorphic, or agree on all invariant structure---is wholly and only captured by $\tilde{\psi}$.  The (salient) sense in which they are different---that they assign different values of the metric to the ``same'' point---is given by an entirely different map, namely $1_M$.  But---and this is the central point---one cannot have it both ways.  Insofar as one wants to claim that these Lorentzian manifolds are physically equivalent, or agree on all observable/physical structure, one has to use $\tilde{\psi}$ to establish a standard of comparison between points.  And relative to this standard, the two Lorentzian manifolds agree on the metric at every point---there is no ambiguity, and no indeterminism.  (This is just what it means to say that they are isometric.)  Meanwhile, insofar as one wants to claim that these Lorentzian manifolds assign \emph{different} values of the metric to each point, one must use a different standard of comparison.  And relative to \emph{this} standard---that given by $1_M$---the two Lorentzian manifolds are \emph{not} equivalent.  One way or the other, the Hole Argument seems to be blocked.\footnote{\citet{Iftime+Stachel} also suggest that the Hole Argument can be blocked---but only if one adopts a different formalism for relativity theory, where one effectively takes the models of the theory and quotients out by isometry.  My view here is different: I am claiming that the Hole Argument is blocked with \emph{no} modification to the formalism of relativity theory.  One only has to recognize the mathematical significance of an isomorphism.}

To close this section, I will address a worry that some readers might have at this stage.  The worry is that the diffeomorphism associated with the Hole Argument is meant to be a so-called ``active'' diffeomorphism, whereas I am interpreting it as a ``passive'' diffeomorphism.\footnote{This distinction is described, for instance, by \citet[pgs. 438--9]{Wald} and \citet[Supplement on Active and Passive Covariance]{NortonSEP}.}  Actually, it seems to me that this distinction is itself symptomatic of the problem with the argument.  The distinction between ``passive'' and ``active'' diffeomorphisms in the present sense is supposed to arise in the context of the action of an induced pushforward map on some field.  The ``passive'' interpretation of such a map is supposed to correspond to a (mere) re-labeling of points, effectively leaving the fields unchanged.  This, I claim, is the only way in which $\tilde{\psi}$ can be interpreted---and, for that matter, how $\tilde{\varphi}$ in the previous section should be interpreted as well.

The ``active'' interpretation, meanwhile, involves keeping the points unchanged (i.e., by somehow fixing the labels of the points, perhaps with a local coordinate system) and assigning fields differently to each point.  I will not dispute that there are contexts in which an ``active'' diffeomorphism may make sense---for instance, when one considers automorphisms of a manifold or a Lorentzian manifold and one has the identity map to refer to as a background standard of ``sameness'' of points.\footnote{The (local) isometries generated by Killing fields are an example of this sort of automorphism.}  But one has to be very careful.  In the general case, where an isometry is \emph{not} an automorphism, one simply does not have the structure to make sense of ``active'' diffeomorphisms as realizing some sort of equivalence.  As I have argued, any claim of equivalence must be relativized to a choice of isometry, and if the identity map does not realize an isometry, then the identity map plays no role in the equivalence.  In particular, the relevant standard of ``same'' labeling available is the one determined by the choice of isometry.  When one ``fixes'' a labeling system and then considers how fields transform within that system, one is conflating the identity map with the (non-trivial) diffeomorphism generating the pushforward map, just as in the Hole Argument.

This last point can be made in fully physical terms.  Suppose the ``fixed'' labeling system represents coordinates associated with an observer's laboratory instruments.  Then holding this constant while one transforms, say, the metric by the pushforward map of a non-trivial diffeomorphism will \emph{not} give rise to a physically equivalent configuration by the lights of this observer.  According to the observer, all of her stuff has been moved around!  This is despite the fact that the two configurations are isometric, in the sense that there \emph{exists} an isometry relating them.

\section{An Argument from Classical Spacetime Theory}

To see what is going on in the previous section, it may be helpful to draw a contrast with a (superficially) similar argument in the context of classical spacetime theory.\footnote{\citet[\S 7]{PooleyOHPP} also contrasts the Hole Argument with the argument I describe here.  As he draws it, however, the distinction concerns whether empirically indistinguishable models of a theory have qualitative differences (according to that theory). This is closely related to, but not quite the same as, the point I am trying to make.  My point is that even adherents to a strong form of substantivalism about classical spacetime structure can still generate notational variants, and these notational variants play no role in the standard (convincing) argument for adopting Galilean spacetime (say) over Aristotelian or Newtonian spacetime.}   Consider the geometric structure sometimes known as ``Newtonian spacetime''.  This can be described as an ordered quintuple $(M,t_{ab},h^{ab},\nabla,\xi^a)$, where $M$ is a smooth four dimensional manifold, $t_{ab}$ and $h^{ab}$ are classical temporal and spatial metrics, respectively, $\nabla$ is a smooth, flat, geodesically complete derivative operator compatible with the classical metrics, and $\xi^a$ is a smooth timelike vector field satisfying $\nabla_a\xi^b=\mathbf{0}$.\footnote{This is not the occasion for a detailed treatment of classical spacetime structure.  For more on this topic, see \citet[Ch. 2]{EarmanWEST} and \citet{MalamentGR}.  (Malament does not describe Newtonian spacetime, but what I am calling Newtonian spacetime is what he calls classical spacetime, with the additional structure of a standard of rest, given by $\xi^a$.)}  A Newtonian spacetime in this sense is invariant under any diffeomorphism generated by a constant timelike or spacelike vector field, in the sense that given such a diffeomorphism, $\vartheta:M\rightarrow M$, we find $(M,t_{ab},h^{ab},\nabla,\xi^a)=(M,\vartheta_*(t_{ab}),\vartheta_*(h^{ab}),\vartheta_*(\nabla),\vartheta_*(\xi^a))$.  These diffeomorphisms---or more precisely, the maps $\tilde{\vartheta}$ between Newtonian spacetimes generated by these diffeomorphisms---are naturally interpreted as representing temporal and spatial translations (or a combination of the two).\footnote{Note that there is a certain poignance here, in that what we now call ``Newtonian spacetime'' is actually invariant under the classical example of a ``Leibniz shift'', namely a transformation that moves everything to the left by five feet.  One might take this as evidence that ``Newtonian spacetime'' is not what Leibniz and Clarke had in mind.}  Note, importantly, that here the automorphism from $M$ to itself, $\vartheta$, gives rise to an automorphism of the Newtonian spacetime structure.  In this sense, $\vartheta$ (or really, $\tilde{\vartheta}$) is a symmetry of Newtonian spacetime.

Now that I have defined this spacetime, I am going to describe three ways of modifying it.  First, consider what happens if we add matter to a Newtonian spacetime.  We can represent matter in this context by a smooth scalar field, $\rho$; we can then write the Newtonian spacetime with this matter field as $(M,t_{ab},h^{ab},\nabla,\xi^a,\rho)$.  Assume that $\rho$ has compact spatial support, like a beach ball.  (So, the support of $\rho$ looks like a tube through four dimensional spacetime.)  Such a matter distribution is manifestly \emph{not} invariant under spatial translations, even though the spacetime itself is.  This can be made precise by noting that if $\vartheta:M\rightarrow M$ is a diffeomorphism generated by a constant (non-vanishing) spacelike vector field, then $\vartheta$ does \emph{not} give rise to an automorphism of $(M,t_{ab},h^{ab},\nabla,\xi^a,\rho)$, even though it does give rise to an automorphism of $(M,t_{ab},h^{ab},\nabla,\xi^a)$.  Nonetheless, $\vartheta$ may still be used to generate an isomorphism of this new Newtonian spacetime+matter structure, by defining a new matter field, $\rho\circ\vartheta^{-1}$.  Now the two Newtonian spacetime+matter structures, $(M,t_{ab},h^{ab},\nabla,\xi^a,\rho)$ and $(M,t_{ab},h^{ab},\nabla,\xi^a,\rho\circ\vartheta^{-1})$, are isomorphic, with the isomorphism given by a map generated by $\vartheta$.  These two structures are fully equivalent; they both represent (or rather, have the capacity to represent) the same beach ball.  In particular, there is no ambiguity regarding where the beach ball is located, because one only gets this sense of equivalence by associating points using $\vartheta$.

Now consider what happens if we add further structure to the Newtonian spacetime.  One can do this by picking out a particular maximal integral curve of $\xi^a$, $\gamma:\mathbb{R}\rightarrow M$.  The resulting spacetime structure, $(M,t_{ab},h^{ab},\nabla,\xi^a,\gamma)$, is sometimes called ``Aristotelian spacetime''.  It differs from Newtonian spacetime in that, in addition to a privileged standard of rest, there is now a privileged center of the universe.  Much like the beach ball described above, this new structure is not invariant under spatial translations---in other words, the diffeomorphism $\vartheta$ above does not give rise to an automorphism of $(M,t_{ab},h^{ab},\nabla,\xi^a,\gamma)$.  Still, it \emph{does} generate an isomorphism, now between the Aristotelian spacetimes $(M,t_{ab},h^{ab},\nabla,\xi^a,\gamma)$ and $(M,t_{ab},h^{ab},\nabla,\xi^a,\vartheta\circ\gamma)$.  In other words, these two Aristotelian spacetimes represent (or, again, have the capacity to represent) the exact same physical situations---they even agree on where the center of the universe is at all times---even though they are related by a map generated by a diffeomorphism that, in a different context, could be interpreted as a spatial translation.  How could this be?  It is because these two spacetimes are only isomorphic, equivalent, etc. insofar as one associates points using the map $\vartheta$.  By the standard according to which they are isomorphic, they agree on all structure.

As a final variation on the theme, consider what happens when we put the beach ball, represented by $\rho$, into Aristotelian spacetime.  We now get an Aristotelian spacetime+matter structure of the form $(M,t_{ab},h^{ab},\nabla,\xi^a,\gamma,\rho)$.  This may be a perfectly adequate representation of the beach ball in spacetime.  But suppose that, instead of using $\rho$, we represented the beach ball using $\rho\circ\vartheta^{-1}$, yielding $(M,t_{ab},h^{ab},\nabla,\xi^a,\gamma,\rho\circ\vartheta^{-1})$.  One might then argue as follows: $(M,t_{ab},h^{ab},\nabla,\xi^a,\gamma,\rho\circ\vartheta^{-1})$ \emph{also} provides a perfectly adequate representation of the beach ball in spacetime.  The reason is that these two structures disagree only with regard to where the beach ball is located relative to the center of the universe, and since the center of the universe is unobservable, even in principle, the two Aristotelian spacetime+matter structures are physically, or at least empirically, equivalent.  But now we have a problem: these two Aristotelian spacetime+matter structures are \emph{not} isomorphic relative to $\vartheta$ or any other map.\footnote{One has to be careful here, since $\vartheta$ \emph{can} be used to generate an isomorphism between two Aristotelian spacetime+matter structures: namely, $(M,t_{ab},h^{ab},\nabla,\xi^a,\gamma,\rho)$ and $(M,t_{ab},h^{ab},\nabla,\xi^a,\vartheta\circ\gamma,\rho\circ\vartheta^{-1})$.  These would be perfectly equivalent for the Aristotelian, relative to the map induced by $\vartheta$.  The example in the text does not concern this transformation, but instead what happens if we do not transform $\gamma$ with $\vartheta$.}

The adherent to Aristotelian spacetime theory is thus forced to either accept that there are physically distinct configurations that nonetheless agree on all observable properties, or else accept---as a substantive additional assumption---that Aristotelian spacetimes differing only in the position of a beach ball relative to the center of the universe must be physically equivalent.  And taking the second of these routes amounts to adopting Newtonian spacetime structure over Aristotelian spacetime structure.

Arguments such as this last one have been very persuasive against adopting Aristotelian spacetime.  And analogous arguments have weighed against Newtonian spacetime, in favor of ``Galilean spacetime'', which lacks a standard of absolute rest (i.e., there is no privileged vector field $\xi^a$).\footnote{I focus on Newtonian and Aristotelian spacetimes here because the automorphisms of Galilean spacetime that fail to be automorphisms of Newtonian spacetime are more complicated to describe in terms of their generating vectors.  But the argument has precisely the same form.}  The Hole Argument, on its face, may appear to be like this argument against Aristotelian spacetime.  But in fact it is different.  Mathematically, the Hole Argument trades on a construction analogous to the first two variations on Newtonian spacetime, where one used an automorphism of one object (in these cases, Newtonian spacetime; in the Hole Argument, the bare manifold) to generate an isomorphism between two other objects.

The crucial observation, here, is that in these three cases, one already has a perfectly good sense in which the structures in question are ``the same''---namely, the sense given by the maps generated by $\vartheta$ (or, respectively, $\tilde{\psi}$ in the Hole Argument above).  One does not need to accept some new form of equivalence in order to avoid an ambiguity regarding which of various possible spacetime structures represents the world, because the relevant standard of equivalence is already manifest in the map that relates the structures in the first place.  It is only in a case where the empirically indistinguishable structures are \emph{not} isomorphic---as in the argument just above regarding Aristotelian spacetime---that one runs into a genuine problem.  As we have seen, however, this is not the situation with the Hole Argument.

\section{The Hole Argument Revisited\label{revisited}}

At this point, the principal arguments of the paper are complete.  But I have said almost nothing about how these arguments relate to the standard interpretation of, or the extensive literature on, the Hole Argument.  I will conclude by turning to this topic.  Even so, this will not be the occasion for a general discussion of the substantivalism/relationism debate.  I will focus on two connections to the rest of the Hole Argument literature.  The first concerns Earman and Norton's own views on the significance of the Hole Argument, as expressed in \citet{Earman+Norton}, and the second concerns the relationship between the present view and other responses to the Hole Argument.

To begin, recall that the moral of the Hole Argument is supposed to be that one must accept that ``diffeomorphic models represent the same physical situation.''\footnote{\label{diffeomorphism} As I indicated in fn. \ref{confusion}, I believe this way of putting the principle is potentially confusing.  The reason is that, as Earman and Norton characterize them, models of a ``local spacetime theory'' consist in a specification of a manifold and a collection of tensor fields on the manifold; ``diffeomorphic models,'' then, are models where the two manifolds are related by a diffeomorphism \emph{and} the associated tensor fields are related by the pushforward map associated with that diffeomorphism.  And here is where the slippage arises.  A diffeomorphism is, unambiguously, an isomorphism of manifolds.  The map relating two ``diffeomorphic models'' in the Earman-Norton sense is not a map (isomorphism or otherwise) between two manifolds.  It is a map between two manifolds endowed with additional structure that is, in a precise sense, \emph{generated} by a diffeomorphism.  The difference matters.  The diffeomorphism in the Hole Argument is an automorphism of the manifold, but not an automorphism of the manifold with additional fields.  My main point in sections \ref{warm-up} and \ref{main-event} has been that running these two kinds of map together is what makes one slip into thinking that the identity map on the manifold has some role to play.}  If one were to modify this slightly, and instead say that the moral is that \emph{isometric} models represent the same physical situations, or even better, isometric models \emph{may be used to} or \emph{have the capacity to} represent the same situations, then the moral is untouched by the considerations in sections \ref{warm-up} and \ref{main-event}.  Indeed, this conclusion follows immediately from the background commitments concerning mathematics and its application to physics that I described in section \ref{introduction}.  On the view described there, once one asserts that spacetime is represent by a Lorentzian manifold, one is committed to taking isometric spacetimes to have the capacity to represent the same physical situations, since isometry is the standard of isomorphism given in the mathematical theory of Lorentzian manifolds.  To deny this would be in effect to insist that it is some other structure---one that is not preserved by isometries---that represents spacetime in relativity theory.

But if the Hole Argument merely supports a view one could have had independent reasons to accept, what becomes of the Hole Argument's intended target---that is, the substantivalist who rejects Leibniz equivalence?  It might seem that the arguments above, if correct, would render this substantivalist position incoherent.\footnote{I take it this would be a problem for the views described here, since it would then seem that mathematical considerations would have addressed the interpretive or metaphysical question about possible worlds after all, suggesting that I have bundled in more metaphysics than I acknowledge.}  But this reaction is too quick.  As I anticipated in section \ref{introduction}, the arguments here do not rule out (or even rule on) substantivalism or relationism.  All I have argued here is that, if one adopts the standard formalism of relativity theory, on which one represents spacetime with a Lorentzian manifold, the Hole Argument does not force one to confront a metaphysical dilemma.

This point may be clearer if we reconsider Earman and Norton's substantivalist in terms of the background views on (applied) mathematics described in section \ref{introduction}.  From that perspective, the substantivalist who denies that isometric spacetimes are equivalent must be asserting that spacetime is not or cannot be adequately represented by (precisely) a Lorentzian manifold, at least for some purposes.  So what sort of structure \emph{would} be adequate to represent spacetime?  Earman and Norton appear to suggest that the substantivalist believes spacetime may be represented by just the \emph{manifold}, not the manifold with metric.\footnote{See \citet[pgs. 518--20]{Earman+Norton}; cf. \citet[pgs. 135-6]{Stachel1993}.  Readers may rightly worry that I am being uncharitable to Earman and Norton, here, or misrepresenting their substantivalist.  But bear with me.}  But, at least from the present perspective, this proposal cannot be right, for several reasons.  One problem is that, on a natural understanding of ``structure'' in mathematics, this proposal implies that the substantivalist takes spacetime to have \emph{less} structure than on the alternative view, insofar as manifolds have less structure than pseudo-Riemannian manifolds.  Even more troubling, such a substantivalist would apparently be committed to the radical view that any two spacetimes are equivalent as long as their associated manifolds are diffeomorphic---whether the two agree on other structure, such as fields defined on the manifold, or not.  Indeed, a substantivalist in this sense would not reject Leibniz equivalence as Earman and Norton describe it.

These considerations suggest that Earman and Norton's substantivalist must actually hold the opposite view, that spacetime is properly represented by a Lorentzian manifold \emph{plus} something else, such that the relevant standard of isomorphism would be more restrictive than isometry.  (One might think of this sort of substantivalist as similar to someone who accepts Aristotelian spacetime over Newtonian spacetime.)  On this reading, one might argue that my characterization of the Hole Argument above missed the point: Earman and Norton take as a premise that spacetime has \emph{at least} as much structure as a model of a local spacetime theory---that is, as a Lorentzian manifold---but may, perhaps, also have more structure.  The Hole Argument, then, does have a substantive role to play: it blocks this possibility, by showing that this extra structure is subject to a pernicious form of indeterminism.  But if this is right, it seems to me that simply putting the point in these terms is helpful.  In particular, it shows that the would-be substantivalist, in order to reply effectively to the Hole Argument, needs to stipulate what the additional structure might be and why we should think it matters. And it is difficult to see how this could be done in a mathematically natural or philosophically satisfying way.\footnote{\label{color} Some authors have appeared to suggest that the substantivalist's additional structure consists in ``spacetime points''.  But manifolds (and Lorentzian manifolds) already have points, and so it is not clear what would be added.  On another way of construing things, it would seem that Earman and Norton's substantivalist wants to consider manifolds in which, in addition to the manifold structure, every point has a unique label.  (\citet{Stachel1993} calls this sort of construction a ``color manifold,'' following an example of Riemann's (pg. 139); he, too, takes the Earman-Norton substantivalist to be proposing that we need this sort of extra structure to represent spacetime (pg. 146).)  This would be a mathematical structure for which \emph{only} the identity map is an isomorphism.  One could certainly do physics with such a structure, but it is hard to see what one would gain.  Indeed, as \cite[pg. 141]{Stachel1993} points out, the fact that we use the particular mathematical structures we \emph{do} use is the end result of a long process of developing and interpreting general relativity---including Einstein's own discovery of and reckoning with the Hole Argument.}

All that said, I think it is fair to say that Earman and Norton did not intend their substantivalist foil to hold \emph{either} of these positions---both are merely reconstructions motivated by the background views on mathematics described at the beginning of this paper.  What Earman and Norton appear to have had in mind is a view on which one represents spacetime with a Lorentzian manifold (and no more), such that the manifold is understood to represent spacetime points in some sense prior to or independently of the value of the metric at those points.  Various metric fields, then, would correspond to different assignments of properties to the spacetime points.  The pushforward of the metric along an automorphism of the manifold would represent different property assignments to the same spacetime points, corresponding to \emph{prima facie} different physical situations.  And \emph{this} view, I take it, is incompatible with the background views on mathematics described in section \ref{introduction}.  The problem is essentially the same as for the Hole Argument itself: this view depends on taking the identity map to provide a prior notion of when the points of two Lorentzian manifolds are ``the same''. But once again, it bears emphasizing that the tension concerns how we use mathematics to represent certain physical or metaphysical possibilities, and not whether the views on the metaphysics of space and time that one would like to represent are tenable in the first place.

The discussion here and in section \ref{main-event}, concerning standards of ``sameness'' for spacetime points, may evoke a now-common response to the Hole Argument.  Many commentators have suggested that one can defend substantivalism by denying haecceitism about spacetime points, which amounts to denying that there exist physically possible worlds that differ only with regard to which spacetime points exhibit certain metrical features.\footnote{This sort of response to the Hole Argument is sometimes known as ``sophisticated substantivalism.''  Varieties of it (with a good deal of variation!) have been advocated by, for instance, \citet{Butterfield}, \citet{Maudlin}, \citet{Stachel}, \citet{Brighouse}, \citet{Rynasiewicz}, \citet{Saunders}, and \citet{PooleyOld, PooleyOHPP}; the expression was coined (and the view criticized) by \citet{Belot+Earman}. See \citet[\S 7]{PooleyOHPP} for an extended discussion.}  One version of this view that is particularly similar to what has been defended here is due to \citet{Butterfield} and \citet{Brighouse}.  These authors argue that the appropriate notion of ``sameness'' of points across possible worlds (or spacetimes) should be given by the qualitative properties of the points, including metrical properties.  This leads them to conclude that, given two isometric spacetimes, the right trans-world (or trans-spacetime) identity relation is the one given by the isometry relating them, since this map takes points to points with the same qualitative properties.\footnote{\citet{Butterfield} goes on to argue that given a collection of isometric spacetimes, at most one of them can represent a given physical situation.  \citet{Brighouse}, meanwhile, holds that all of these isometric spacetimes are physically equivalent, in the sense that they can represent precisely the same isometric spacetimes.  In this sense, then, the views of the present paper are somewhat closer to Brighouse than to Butterfield, though most of what I say here applies to both.}

Brighouse puts the point as follows.
\begin{quote}\singlespacing
[The Hole Argument assumes] that a substantivalist identifies spacetime points across the possible worlds in question independently of any of their qualitative properties.  It is assumed that independently of the properties of points coded by the fields $O_i$ the issue of identifying points across possible worlds has been settled.  The substantivalist who then wants to claim that Leibniz equivalent models represent the same world has to deal with a given spacetime point having different properties in worlds that she wants to say are the same.  But realism about spacetime points [i.e., substantivalism], just as realism about any other entities, does not commit you to a transworld identification thesis that is independent of the qualitative properties of an object. \citep[p. 121]{Brighouse}\end{quote}
In other words, Earman and Norton commit their substantivalist to the view that there is a prior notion of identity between points of isometric spacetimes---i.e., that the identity map is privileged.  Without that assumption, the Hole Argument has no force. This response to the Hole Argument, understood as a diagnosis of why the Hole Argument should not be taken to have the force it is often credited with, is very close indeed to the view presented here.
But there are also several differences that are worth emphasizing.  The first difference is that Brighouse and Butterfield both aim to articulate and defend varieties of substantivalism, and their responses to the Hole Argument are cast in those terms.  I do not take myself to be defending substantivalism, or relationism for that matter, here.\footnote{In fact, I am not sure what the difference is supposed to be.  Once substantivalists agree that spacetime is (or can be) represented by precisely a Lorentzian manifold, that isometries are isomorphisms of this structure, and thus that isometric spacetimes are equally good representations of the world---i.e., that they do not correspond to physically different worlds---it is difficult to see how to distinguish the two views.  At least, no physics is at stake.  \citet[pgs. 578-9]{PooleyOHPP} and \citet[\S 3]{Curiel} draw similar conclusions, as do many others.}  In a sense, my goal is to show that one can and should reject the Hole Argument along Butterfield-Brighouse lines, independently of any substantivalist thesis.

A second, related, difference is that Butterfield and Brighouse base their arguments on a view of trans-world object relations motivated by Lewisian counterpart theory.  One gets the impression that to respond to the Hole Argument in the way these authors do, one needs to be committed to Lewis' system.  My arguments here, meanwhile, have required no such commitments.  Of course, one does have to accept the views about (applied) mathematics described in section \ref{introduction}, so I have not gotten something for nothing.  But at very least, the starting point for the present paper is quite different from Butterfield and Brighouse's, and perhaps less controversial.

This difference in starting points highlights a more significant difference between the view defended here and the one Butterfield and Brighouse defend.  Butterfield and Brighouse, and indeed, virtually all of the literature on the Hole Argument and anti-haecceitistic substantivalism, take for granted that there are \emph{mathematical} differences to reckon with between isometric spacetimes---that is, that there are distinct models of relativity theory that differ only with regard to which spacetime points exhibit certain metrical features---and that the puzzle concerns whether these mathematical differences correspond to differences of physical possibility.  In other words, the Butterfield-Brighouse response to the Hole Argument appears to grant that the identity map plays a privileged \emph{mathematical} role in comparing isometric spacetimes.  They then question whether it also plays a privileged physical or metaphysical role in distinguishing possibilities.  The argument here, meanwhile, was that those alleged mathematical haecceities are spurious.

In this regard, more than Butterfield or Brighouse, I am echoing \citet[\S XVI]{Wilson}, who laments a tendency in the Hole Argument (and among philosophers more generally) to ``focus upon the unwanted extrinsic aspects of'' (pg. 238) mathematical constructions.  The mathematical haecceities in question concern precisely those unwanted ``extrinsic'' features of isometric Lorentzian manifolds; ``intrinsically'', isometric Lorentzian manifolds are the same, in the sense given by the isometry.

\section*{Acknowledgments}
This material is based upon work supported by the National Science Foundation under Grant No. 1331126.  Thanks to Steve Awodey, Jeff Barrett, Thomas Barrett, Jeremy Butterfield, Erik Curiel, Ben Feintzeig, Sam Fletcher, Hans Halvorson, Stephan Hartmann, David Malament, John Manchak, Oliver Pooley, Marian Rogers, and Sarita Rosenstock for discussions related to the topics discussed in this paper.  I am especially grateful to Gordon Belot, Carolyn Brighouse, Jeremy Butterfield, Erik Curiel, Ben Feintzeig, Sam Fletcher, Thomas M\/{o}ller-Nielsen, David Malament, John Norton, Brian Pitts, Bryan Roberts, Chris Smeenk, David Wallace, Chris W\"uthrich, and three anonymous referees for their extremely helpful comments on previous drafts.

\singlespacing
\bibliography{holes}
\bibliographystyle{elsarticle-harv}
\end{document}